\documentclass[aps,twocolumn]{revtex4}

\usepackage{graphicx}
\usepackage{color}
\usepackage{topcapt}
\usepackage{booktabs}
\usepackage{amsmath, amsthm, amssymb}

\begin{document}

\bibliographystyle{apsrev}

\title{Characterizing the Gravitational Wave Signature from Cosmic String Cusps }
\author{\surname{Joey} Shapiro Key and {Neil} J. Cornish}

\affiliation{Department of Physics, 
  Montana State University, Bozeman, MT 59717}
\date{\today}

\begin{abstract}

Cosmic strings are predicted to form kinks and cusps that travel along the string at close to the speed of light.
These disturbances are radiated away as highly beamed gravitational waves that produce a burst like pulse as the
cone of emission sweeps past an observer. Gravitational wave detectors such as the Laser Interferometer Space
Antenna (LISA) and the Laser Interferometer Gravitational wave Observatory (LIGO) will be capable of detecting these
bursts for a wide class of string models. Such a detection would illuminate the fields of string theory, cosmology, and
relativity. Here we develop template based Markov Chain Monte Carlo (MCMC) techniques that can efficiently detect and
characterize the signals from cosmic string cusps. We estimate how well the signal parameters can be recovered by
the advanced LIGO-Virgo network and the LISA detector using a combination of MCMC and Fisher matrix techniques. We
also consider {\em joint} detections by the ground and space based instruments. We show that a parallel tempered MCMC
approach can detect and characterize the signals from cosmic string cusps, and we demonstrate the utility of this
approach on simulated data from the third round of Mock LISA Data Challenges (MLDCs).

\end{abstract}

\maketitle

\section{Introduction}

Cosmic strings were first predicted to form during symmetry breaking in the very early universe~\cite{Kibble:1976db}.
Strings with $G\mu \sim 10^{-6}$, where $\mu$ is the string tension and $G$ is Newton's constant, were studied as
possible seeds for structure formation and the source of anisotropies in the cosmic microwave background (CMB)
radiation~\cite{Vilenkin:2000qd}. Observations of the CMB combined with other cosmological
data sets imply that $G\mu < 2.3\times 10^{-7}$, which rules out cosmic strings as the source of structure in
the universe~\cite{Seljak:2006kh}. This bound makes field theoretic cosmic strings poor candidates for detection by
current gravitational wave observatories, though the next generation of detectors will be sensitive to
gravitational waves from field theoretic strings~\cite{Siemens:2007fv}.

Interest in detecting gravitational wave signatures from cosmic string networks has increased following
the realization that cosmic superstrings can be produced in a variety of string inspired inflationary
models~\cite{Sarangi:2002ss,Berezinsky:2001qc,Damour:2000mw,Jones:2002kl,Jackson:2005ff,Jones:2003mi,Copeland:2004pi,Dvali:2004qa}.
Networks of cosmic superstrings can evolve quite diffently than their field theoretic cousins, making them
more promising candidates sources of gravitational waves.
Cosmic superstrings could potentially be observed via gravitational lensing, pulsar timing, observations of the
cosmic microwave background, or from the energy they radiate in the form of gravitational waves~\cite{Vilenkin:2000qd}.
		
A string network consists of long, horizon sized strings and loops that tend to intersect and reconnect, forming discontinuities known as kinks.
Cusps with high Lorentz boosts will also generically form on strings and loops.  A cosmic string network will produce a gravitational wave
background, as well as bursts of gravitational waves from cusps and kinks that stand out above the background. The strongest bursts come from cusps, where small portions of the string will be traveling near the speed of light, leading to the emission of a beam of gravitational radiation~\cite{Damour:2001mz, Damour:2005vn}.  Current and future ground based gravitational wave detectors (LIGO-Virgo-GEO and Advanced LIGO) and the planned NASA/ESA space based detector (LISA) may be able to detect such signals and determine several properties of the string network~\cite{Siemens:2006dp,DePies:2007rq,Siemens:2008fc}.

Here we consider how these cusp signals may be detected, and how well the signal parameters may be inferred using data from ground and space based gravitational wave observatories. The broad spectrum nature of the signals raises the
possibility of joint detection in space and on the ground, and we investigate how this impacts parameter estimation.
To our knowledge this is the first time that joint LIGO-Virgo-LISA observations have been considered for any
gravitational wave source.

The signal parameters that can be measured depend on combinations of the key physical parameters,
so observations of individual bursts are not enough to constrain the string model.  Ultimately it will be quantities such as
the event rate as a function of burst amplitude, and the power spectra of the un-resolved confusion background from more distant bursts and the decay of loops that will provide the strongest model constraints.  In this paper we focus on individual bursts and defer the analysis of what can be learned from a global analysis for a later publication~\cite{Siemens:2008fc}.

It has been pointed out that detecting cosmic strings with a microlensing quasar survey has been essentially ruled out by the low event rates ($ \sim 10^{10}$ quasars sources would have to be monitored for a year to expect a few events) and the long lensing periods (20-40 years) that are predicted~\cite{Kuijken:2008kc}.  Still, loops of cosmic string with tensions in the range $ 10^{-10} < G\mu < 10^{-6}$ are predicted to produce microlensing of stars in the local group of galaxies ~\cite{Chernoff:2007hs}.  A detection of a cosmic string by a gravitational wave detector could thus be followed by an electromagnetic observation by looking in the direction determined by the gravitational wave observation for microlensing events.  This possibility motivates accurate parameter estimation from the gravitational wave detection to provide well determined sky location and orientation information for the string.

The paper is organized as follows: Section \S~\ref{signal} describes the gravitational wave signals from cosmic string cusps
and our method for generating signal templates; Section \S~\ref{mcmc} details the implementation of our Markov Chain Monte
Carlo search technique; Section \S~\ref{mldc} provides an illustration of the effectiveness of our search by applying
it to the simulated data from the Mock LISA Data Challenge; and Section \S~\ref{aligo} considers advanced ground based
detectors and joint LIGO-Virgo-LISA observations of cusp signals.

\section{Gravitational Wave Signature}\label{signal}

The gravitational wave signature from a cosmic string cusp is very simple.  In the limit that the line of sight to the cusp
is coincident with the axis of the emission cone, the waveform is linearly polarized and described by the power law
\begin{equation}\label{cusp}
h(t) = 2 \pi A \vert t - t_* \vert^{1/3} \, .
\end{equation}
Here $t_*$ is the time when the observer sees the intensity peak. Equation (~\ref{cusp}) is a good approximation for
time seperations $\vert t - t_* \vert$ short compared to the light crossing time, $L/c$ of the feature that produces
the cusp~\cite{Siemens:2003sp}.  The overall amplitude $A$ is related to the
distance to the cusp $r$, the string tension $G\mu$, and the characteristic length scale $L$ as
\begin{equation}
A \sim \frac{G\mu L^{2/3}}{r} \, .
\end{equation}
Since $L$ and $r$ are unknown, a measurement of $A$ does not directly reveal the string tension.

The signal is slightly more complicated for off-axis observations.  For small viewing angles $\alpha$, defined as the angle
between the line of sight and the axis of the emission cone, the main effect is to round off the cusp waveform (\ref{cusp}) such that
the power spectrum decays rapidly for frequencies greater than $f_{\rm max} \sim 2/(\alpha^3 L)$. Once again, the
observable quantity, $f_{\rm max}$ does not directly reveal the physical quantity of interest, $L$, since it also
involves the unknown viewing angle $\alpha$.

The observed gravitational wave signal will also depend on how the waveform is projected onto the detector,
which will depend on the sky location ($\theta, \phi$) and the polarization angle $\psi$. Thus, the detected signal depends
on the parameters $A,t_*,f_{\rm max},\theta,\phi,\psi$. These six measurable quantities depend on the two intrinsic source
parameters $(G\mu, L)$ and the six extrinsic (observer dependent) quantities $(r,\alpha, t_*, \theta,\phi,\psi)$. Clearly
the observation of a single burst is insufficient to uniquely determine the intrinsic source parameters. These can only be
determined by considering the full gravitational wave signal from a string network, including the spectrum of the background
and the rate distribution of the brighter bursts~\cite{Siemens:2008fc}.

\subsection{Template Generation}	

Cosmic string cusp templates are easy to generate in the frequency domain as the time dependence of the
detector response can be ignored: The effective duration of the burst
is set by the lowest frequencies the gravitational wave detector can detect. For LISA this is $f_{min} \sim 10^{-5}$ Hz,
and for advanced LIGO the limit is $f_{min} \sim 10$ Hz, which leads to effective durations of $10^5$ seconds and $0.1$ seconds
respectively. In both cases the duration of the burst is short compared to the time scale over which the antenna pattern
varies - a year for LISA and a day for LIGO.

The frequency domain waveform for a burst of gravitational waves from a cosmic string cusp is a power law with
an exponential decay beginning at the maximum frequency given by the viewing angle \cite{Siemens:2003sp}. 

\begin{equation}
h(f)=
\begin{cases} Af^{-\frac{4}{3}}   &    f<f_{max}
\\
Af^{-\frac{4}{3}}e^{1-f/f_{max}}  &   f>f_{max}

\end{cases}
\end{equation}

\begin{figure}[htbp]
   \centering
   \includegraphics[width=3.5in]{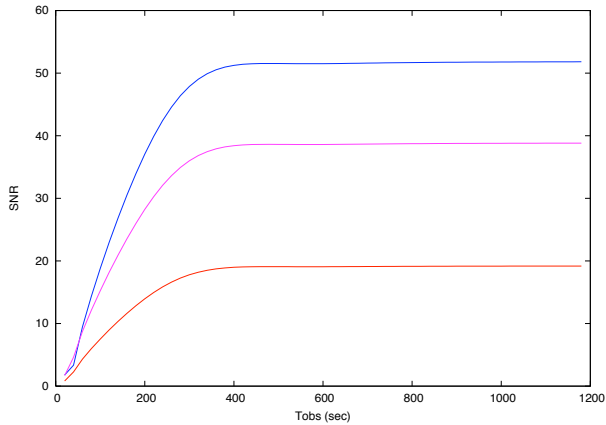}
   \caption{The SNR as a function of observation time for three different sets of parameter values for cosmic string cusp bursts of gravitational waves from the MLDC.}
   \label{fig:SNR}
\end{figure}

When convolved with the instrument response and considered relative to the instrumental noise spectrum
the effective duration of the bursts is even shorter than the simple estimates based on the low frequency limit.  Figure~\ref{fig:SNR}
shows that for LISA observations $99.9+\%$ of the full signal to noise ratio (SNR) is reached for observation times of less than 400 seconds, so the stationary detector approximation is very accurate.  

The SNR for a template $h$ is calculated:
\begin{equation}
\mbox{SNR}=(h|h)^{1/2}
\end{equation}
where the noise weighted inner product for the independent data channels $\beta$ over some observation time $T_{obs}$ is defined as
\begin{equation}
(a|b) = \frac{2}{T_{obs}}\sum_{\beta}\sum_{f}\frac{a_{\beta}^{*}(f)b_{\beta}(f) + a_{\beta}(f)b_{\beta}^{*}(f)}{S_n^{\beta}(f)}
\end{equation}
and $S_n(f)$ is the one-sided noise spectral density in each channel.

\subsubsection{LISA Instrument Response}

We adopt the standard MLDC ecliptic coordinate system with origin at the barycenter. 
The individual data streams from the six LISA phase meters can be combined to cancel out the laser phase noise and form
Time Delay Interferometry (TDI) variables~\cite{Estabrook:2000ef}. The data sets for MLDC Challenge 3.4 included
both the phase meter outputs and the complete set of Michelson style TDI variable $\{X,Y,Z\}$. The latter can be used to
construct three noise orthogonal data streams that are similar to the $\{A,E,T\}$ variables described in
Ref.~\cite{Prince:2002bx}. Each data stream $s$ contains the response to the gravitational wave signal and additive,
stationary, Gaussian distributed instrument noise: $s_\beta = h_\beta + n_\beta$.

The gravitational wave response is computed by convolving $h(\vec{x}, f)$ with the static limit of the LISA instrument
response~\cite{Cornish:2003fr}. The various time delays in the response appear as phase shifts
that give rise to transfer functions of the form
\begin{eqnarray*}
T_{ij}(f) =  \frac{1}{2}{\rm sinc}[\frac{f}{2f_*}(1-\hat{k}\cdot\hat{r}_{ij})] \exp{\{-i[\frac{f}{2f_*}(3+\hat{k}\cdot\hat{r}_{ij})]\}} \\
 +  \frac{1}{2}{\rm sinc}[\frac{f}{2f_*}(1+\hat{k}\cdot\hat{r}_{ij})] \exp{\{-i[\frac{f}{2f_*}(1+\hat{k}\cdot\hat{r}_{ij})]\}}
\end{eqnarray*}
where $\hat{k}$ is the propagation direction of the gravitational wave, $\hat{r}_{ij}$ is the unit vector pointing
from spacecraft $i$ to spacecraft $j$, and $f_*=1/(2\pi L)$ is the LISA transfer frequency for arm length $L$.
The transfer function has zeros at frequencies that depend on the propagation direction. This gives LISA angular resolution
for burst sources since bursts from different sky locations $(\theta,\phi)$ will project differently onto the detector.

\begin{figure}[htbp]
   \centering
   \includegraphics[width=3.5in]{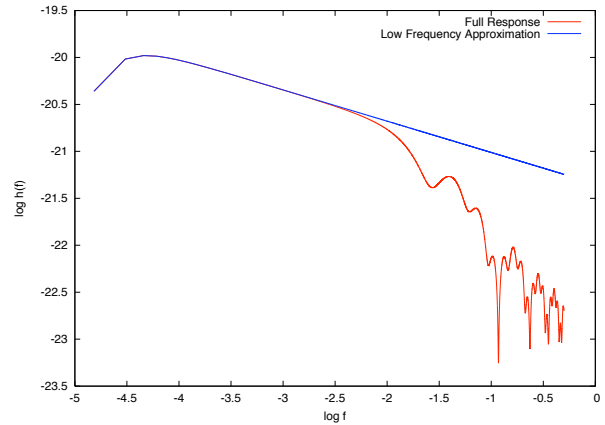}
   \caption{The low frequency approximation to the LISA response compared to the full response for the same source.}
   \label{fig:lowf}
\end{figure}

It is the high frequency component of the gravitational wave signal from a cosmic string cusp that determines
the resolution of the source on the sky.  Cusp seen with large viewing angles result in signals with frequency
cut-offs below the transfer frequency, and as a result, little or no sky resolution.  Figure \ref{fig:lowf}
shows the LISA instrument response in the low frequency limit compared to the full LISA response to the same source. 
The specific location of the minima of the full response depend on the sky location of the source and angular resolution
is achieved because templates with different sky locations have different response shapes.

\subsubsection{LIGO-Virgo Instrument Response}

The LIGO and Virgo instrument response is similar to the low frequency limit of the LISA response, save for a factor of
$\sqrt{3}$ because of the equilateral triangle configuration of LISA as opposed to the $90^{\circ}$ orientation of the
ground based detector arms.

The orientation and location of each detector is most easily defined in an Earth fixed coordinate system with origin
at the geocenter~\cite{Anderson:2001if}. The Earth can be considered stationary for the short duration that
the cusp signal is in the LIGO-Virgo band. This allows us to generate the waveform templates in the frequency
domain with the different times of arrival at each detector appearing as relative phase shifts in the signal.

During the epoch of joint LISA-LIGO-Virgo observations the ground based detectors are expected to be operating in
advanced configurations, so the baseline advanced LIGO wideband noise curve was used for the terrestrial network.

\section{MCMC Techniques}\label{mcmc}

\subsection{Markov Chain Monte Carlo}

The Markov Chain Monte Carlo search method provides a powerful tool for searching large parameter spaces
without creating a large grid of templates~\cite{Metropolis:1953ne,Hastings:1970gb}.  

Instead of stepping along a grid to find the best fit to the data, a template $h(\vec{x})$ is generated at some place
in parameter space, $\vec{x}$, and compared to the data to form the likelihood  $\mathcal{L}(\vec{x})$.
Each of the parameters is then varied by some amount to reach a new place in parameter space, $\vec{y}$.
The quality of the fit at each location is evaluated and the Hastings Ratio $H$ is then formed to compare the
two templates~\cite{Cornish:2005ul}.
\begin{equation}
H=\frac{\Pi(\vec{y})\mathcal{L}(\vec{y})q(\vec{x}|\vec{y})}{\Pi(\vec{x})\mathcal{L}(\vec{x})q(\vec{y}|\vec{x})} 
\end{equation}
\begin{equation}
\mathcal{L}(\vec{\lambda})=e^{-\frac{1}{2}(s-h(\vec{\lambda})|s-h(\vec{\lambda}))} \, . 
\end{equation}
	
The Hastings ratio is built with $\Pi(\vec{x})$, the prior distribution for the template parameters, and $q(\vec{x}|\vec{y})$, the proposal distribution, which is the function that generates proposals for moves from $\vec{x}$ to $\vec{y}$.  Once a jump is proposed, the parameters $\vec{y}$ are adopted with probability $\alpha = {\rm min}[1,H]$.  When $\vec{y}$ is a better fit to the data than $\vec{x}$, $H$ is greater than 1 and the parameters $\vec{y}$ are adopted.  If $\vec{y}$ is a worse fit to the data, $H$ is less than 1, and there is a probability equal to $H$ that the parameters $\vec{y}$ will be adopted.  This process is repeated until some convergence criteria is reached, and the parameter values at each iteration are stored to produce a useful map of the parameter space.  

A properly constructed MCMC search will sample the parameter space with the number of iterations spent at each parameter value proportional to how well that value fits the data.  The chain of values for each parameter is used to form histograms that represent the marginalized posterior distribution functions (PDFs) that show the resolution and expected error for the parameter.

The signal was parameterized by $\vec{x} \rightarrow \{ \ln A, t_*, \ln f_{\rm max},\theta,\phi,\psi\}$, and the priors were
taken to be uniform in these quantities, save for $\theta$, where a uniform sky distributions is given by
$\Pi(\theta)=\frac{1}{2}\sin(\theta)$.

\subsection{Proposal Distributions}

An MCMC search is guaranteed to eventually converge on the posterior distribution regardless of choice of proposal distribution.
In practice the performance of the algorithm is quite sensitive to the choice of proposal distribution, and care must be
taken to ensure that the chains do not get stuck on local maxima of the PDF. We employed several techniques to ensure rapid
exploration of the full parameter space: local coordinate transformations to uncouple the parameters; moves that exploit symmetries
of the likelihood surface to encourage jumps between local maxima; and parallel tempering to encourage wide exploration of the
posterior.

\subsubsection{Gaussian Proposals}

The Fisher Information Matrix $\Gamma$ can be used to propose a jump to parameter values $\vec{y}$ based on the Gaussian approximation to
the Likelihood at $\vec{x}$.  These proposals efficiently explore local maxima by calculating the eigenvectors and
eigenvalues from the Fisher Matrix to propose appropriate sized jumps along independent eigen-directions rather
than the correlated coordinate directions.  

The elements of the Fisher matrix are given by
\begin{equation}
\Gamma_{ij}(\vec{\lambda}) = (h_{,i}|h_{,j})
\end{equation}
where $h_{,i} = \partial_{\lambda_i}  h$.

A Gaussian Fisher proposal distribution is good for local exploration of a parameter space, but more global proposal
distributions are needed to ensure full coverage of the parameter space. 

\subsubsection{Detector Symmetry Based Proposals}

\begin{figure}[htbp]
   \centering
   \includegraphics[width=2in]{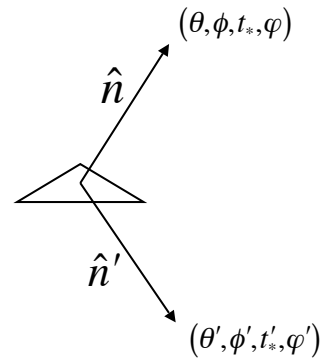}
   \caption{Reflection in the plane of the detector gives a degenerate set of parameter values.  Rotations in the
plane of the detector give two sets of secondary maxima.}
   \label{fig:symmetry}
\end{figure}

Understanding the symmetries associated with the detection of bursts provides another effective proposal distribution.
Any burst short enough to treat the LISA detector as stationary will have a degeneracy such that sky locations related
by a reflection in the plane of the detector with produce an {\em identical} response. There are additional symmetries
in the low frequency limit that result from $120^\circ$ rotations in the plane of the detector. These symmetries are
broken at higher frequencies by the slightly different arrival times of the gravitational waves across the LISA array.
Rotations thus produce two sets of secondary maxima (Figure \ref{fig:symmetry}).

The MCMC searches for cosmic string cusps include proposal jumps from $\vec{x}$ to the set of parameters $\vec{y}$ that
give the degenerate detector response. The mapping between these sets of parameters involves the sky location,
polarization angle, and the time of arrival at the solar barycenter.  The symmetry reflection and rotations are included as possible jumps, but the rotations are rarely accepted unless the chain happens to be at a secondary maxima and the rotation takes the chain to
one of the two primary maxima.

\begin{figure}[htbp]
   \centering
   \includegraphics[width=3in]{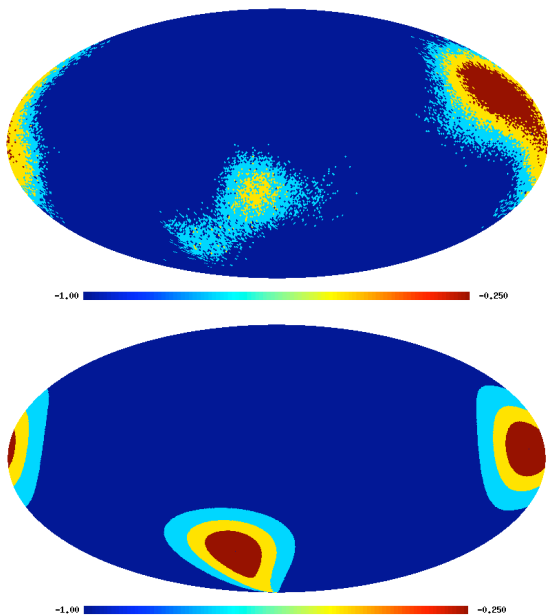}
   \caption{A sky location histogram of a Parallel Tempered MCMC search for MLDC source 3.4.2 on the full sky with quartile contours and
a Fisher matrix approximation for the source location below for comparison (marginalized over the other source parameters).
The all-sky figures use the HEALPix pixelization of the sky (http://healpix.jpl.nasa.gov). }
   \label{fig:source2}
\end{figure}

\begin{figure}[htbp]
   \centering
   \includegraphics[width=3in]{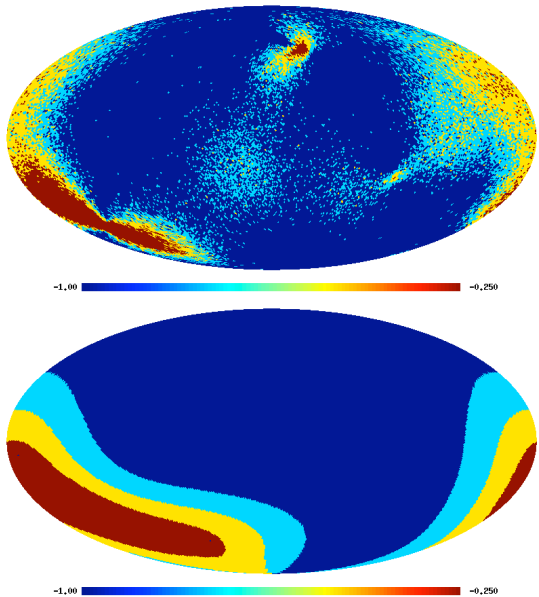}
   \caption{A sky location histogram of a  Parallel Tempered  MCMC search for MLDC source 3.4.4 on the full sky with quartile contours and
a Fisher matrix approximation for the source location below for comparison (marginalized over the other source parameters).}
   \label{fig:source4}
\end{figure}

Our Fisher matrix prediction uses the LISA symmetry and calculates the approximation to the response at the two degenerate sky locations.  While the
likelihood is identical at the two locations in parameter space, the curvature of the likelihood surface is different. The Fisher matrix is useful for
driving jumps in the MCMC searches, but it is not a perfect prediction of the PDF for the
source.  The maps shown in Figures \ref{fig:source2} and \ref{fig:source4} compare the marginalized PDFs for the sky location derived from the Fisher
matrix approximation and from Markov Chain Monte Carlo explorations of two bursts in the Mock LISA Data Challenge 3.4 training data.

We compared MCMC runs with and without the symmetry based jumps, and while parallel tempering did allow the chains to
transition between the various maxima, the mixing was greatly enhanced by including the symmetry based jumps.
A chain with such symmetry jumps moves freely between the two degenerate locations in parameter space, exploring both
peaks and producing bimodal histograms for each source parameter.

The use of a variety of proposal distributions is essential for an efficient search of the parameter space.
In the case of cosmic string cusp gravitational wave sources the symmetry considerations are especially important
to avoid getting stuck on a single maxima.  The detection of a cosmic string cusp by LISA must include the two
degenerate answers, with the possibility of the degeneracy being broken by a simultaneous detection by the ground
based gravitational wave detector network.  

\subsection{Parallel Tempering}

Global exploration of the parameter space is enhanced by creating a set of parallel chains with likelihood surfaces
at different ``Temperatures'' $T$ such that
\begin{equation}
\mathcal{L}_{i}(\vec{x})=\mathcal{L}(\vec{x})^{1/{T_i}} \ \ .
\end{equation}
Chains that explore surfaces with $T\gg 1$ tend to take bigger steps since the contrast between maxima and minima is
decreased, and this encourages wider exploration of the parameter space~\cite{Gregory:2005aw,Gregory:2005dv}. The
chains can exchange parameters according to the Hastings ratio 
\begin{equation}
H_{PT}=\frac{\mathcal{L}_a(\vec{x}_b)\mathcal{L}_b(\vec{x}_a)}{\mathcal{L}_a(\vec{x}_a)\mathcal{L}_b(\vec{x}_b)} \ \ ,
\end{equation}
for chains with temperature $T_a$ and $T_b$ and parameters $\vec{x}_a$ and $\vec{x}_b$, respectively. 

We implement the parallel tempering method for $N_C$ chains with the $T$ values given by
\begin{equation}
T_{i}=(\Delta T)^{i-1}
\end{equation}
where
\begin{equation}
\Delta T=(T_{\rm max})^{\frac{1}{N_{C}-1}} \ \ .
\end{equation}

Each chain explores its own likelihood surface until an exchange of parameters with a neighboring chain is proposed
as a step in the MCMC algorithm.  If the proposal is accepted, the two chains trade parameter values and continue to
explore from this new location.  Ideally, the chain with the highest $T$ value has an effective likelihood surface that
is smooth enough that the chain makes bold jumps and freely explores the entire range of parameters.  The discovery of a
favorable location in parameter space is then propagated down to the $T=1$ chain as exchange proposal jumps are accepted
by adjacent chains.  Only the $T = 1$ chain samples the true PDF and is used to produce the parameter histograms.

There are several trade offs that go into the choice of $\Delta T$ and $N_C$. The hottest chain should have
an effective signal-to-noise ratio ${\rm SNR}_{\rm eff} = {\rm SNR}\, T_{\rm max}^{-1/2} < 5$ to ensure complete
coverage of the parameter space, and the temperature increment between chains should not be much larger
than $\Delta T < 2$ to ensure good mixing between chains. For example, to explore a signal with ${\rm SNR} = 30$,
a good choice of parameters would be $\Delta T = 1.4$ and $N_C = 12$. While the cost per iteration of a parallel tempered MCMC
algorithm is $N_C$ times larger than a standard MCMC algorithm, the improvement in the mixing, especially between widely
separated maxima, ultimately leads to a far more efficient exploration of the posterior.

\section{The Mock LISA Data Challenges}\label{mldc}

To foster the development of LISA data analysis techniques, and to demonstrate mission readiness,
a series of Mock LISA Data Challenges has been conducted~\cite{Babak:2008tk}.  The simulated data sets include
anticipated sources of instrument noise and a wide range of astrophysical sources.  Challenge 3.4 contains gravitational
waveforms for burst signals from cosmic string cusps.  Data analysis techniques can be used to separate the signals from
the noise and estimate the parameters of the sources.  The MLDC Taskforce has created both training data sets with an answer
key for the injected source parameters and blind data sets with a secret answer key.  The injected parameter values for the blind
data sets are revealed after a set deadline and submissions by data analysis groups are evaluated for accuracy~\cite{Babak:2008pd}.  

Challenge 3.4 is comprised of a month long data set ($2^{21}$ samples with 1 second sampling) with cosmic string cusp waveforms
injected with a Poisson event rate of five events per month.  The training data set includes an answer key with the number of
injected sources and their parameter values.  The training data happens to contain five sources, but there is a good chance
that the blind data could have between two and eight sources.  The MLDC sources have SNR uniformly distributed in [10, 100]
and $\log(f_{\rm max}/{\rm Hz})$ uniformly distributed in [-3,1].  This is also the first MLDC data set with non-symmetric instrument noise.
The cusp burst sources can be found using a symmetric approximation for the noise (leading to a small systematic bias
in the recovered parameters), or the source parameters and the individual noise levels can be fitted simultaneously in the search.

The time of arrival at the solar system barycenter ($t_{*}$) is highly correlated with the sky location of the source.
A search for $t_{*}$ leads to poor determination of the time of arrival of the burst due to the inherently poor resolution
of the sky location.  A better choice of variable is the time of arrival at the guiding center of the LISA constellation
($t_{\triangle}$).  The detector time of arrival is not as correlated with the sky location parameters, resulting in
better conditioned Fisher Information Matrices to drive the local jumps of the Markov Chain.

\subsection{Training Data Results}

The training data was analyzed without reference to the answer key so as to mimic the steps that will be taken to analyze
the blind challenge data. The parallel tempering technique takes care of both detection and characterization, so the analysis
does not have to be be broken up into distinct stages. On the other hand, running the search on the full $\sim 2\times 10^6$ seconds
of data to find signals with duration $\sim 10^{3}$ seconds is not very efficient, so we adopted the strategy of dividing the
full data set into 64 segments of length 32,768 seconds. Time domain filters were used to limit spectral leakage, and the
finite response of these filters meant that signals in the first and last $\sim 10\%$ of each segment had to be discarded.
To ensure full coverage, a second pass was performed using segments offset from the first by $16,384$ seconds.

The basic parallel tempered MCMC algorithm was able to both detect and characterize the cosmic string cusp signals in each
segment. The ``burn-in'' time for the chains to lock onto the signals was shortened by several orders of magnitude
by analytically maximizing over the time of arrival at the detector center and the amplitude. These maximizations render the
chains non-Markovian, and must be turned off after the burn-in is complete, and the samples from the burn-in must be discarded.

A fully-fledged Bayesian MCMC analysis of the data would involve conditions to decide when the burn-in phase was complete, the
ability to search for multiple cusp signals simultaneously, and evidence based selection of the putative detections.
While parallel tempered MCMC algorithms can do all of these things, we settled on a less sophisticated approach that could
be implemented with less effort. The first stage of the analysis was to search each data segment using $N_C=12$ chains with
$\Delta T = 1.55$ and $N=10,000$ iterations. A simple SNR threshold was used to decide if a source had been
found in the data segment. Triggers with ${\rm SNR} = (s\vert h)^{1/2} > 8$ were recorded for further analysis
(the loudest noise triggers had ${\rm SNR} < 6$). If a trigger was found the signal was regressed from the data and the
search repeated (in other words the search is sequential rather than simultaneous).

The initial search did not fit for instrument noise levels, but found all five signals in the training data and recovered the
source parameters to good accuracy (Table I).  Since the segmented data is searched twice, we expect to find each source twice,
but one trigger happened to fall near the boundary between segments and was thus discarded.  The source was found on the offset
pass. 

\begin{table}[htbp]
   \centering
   \begin{tabular}{|c|c|c|c|c|}
      \toprule
      \multicolumn{2}{c}{} \\
      \hline
 Source & $f_{\rm max}$ (Hz)   & SNR   &  $\Delta f $ (Hz) & $\Delta t$ (sec) \\ 
	\hline

      \midrule
        
3.4.0 & 2.36e-3 & 53.54   & 3.98e-5   & 4.64            \\
      &         & 53.99   & 7.70e-5   & 5.42           \\
	\hline
3.4.1 & 1.15    & 21.46    & 1.11   & 2.54             \\
      &         & 21.27    & 2.62   & 2.94             \\ 
	\hline
3.4.2 & 0.46    & 31.07    & 0.329   & 0.15           \\
	\hline
3.4.3 & 1.15e-2 & 73.87    & 3.82e-4   & 1.16            \\
      &         & 76.58    & 9.11e-4   & 0.38         \\      
\hline
3.4.4 & 2.27    & 14.07    & 2.25   & 3.44             \\
      &         & 14.12    & 4.14   & 2.45         \\   
 \hline
      \bottomrule
   \end{tabular}
   \caption{The triggers produced by a search of the MLDC 3.4 training data. All but one of the sources
was detected in both passes through the data (one signal happened to straddle a data segment boundary).
The injected $f_{\rm max}$ values are listed, along with
the recovered values for SNR.  The difference between the injected and recovered parameter values for
$f_{max}$ and $t_\triangle$ are given in the last two columns. Note that sources with $f_{\rm max} > f_*$
have large $\Delta f$'s. These parameter errors are from the first stage of the filtering - the
fit improves after the second stage.}
   \label{tab:booktabs}
\end{table}

The initial detection of the burst could be done at similar cost using the template grid approach used to search for
cosmic string bursts in LIGO data~\cite{Siemens:2006dp}. The main value of the MCMC approach is that it allows us to construct
the joint posterior distribution function during the parameter estimation stage of the analysis.

\subsection{Parameter Estimation}

\begin{figure}[htbp]
   \centering
   \includegraphics[width=3in]{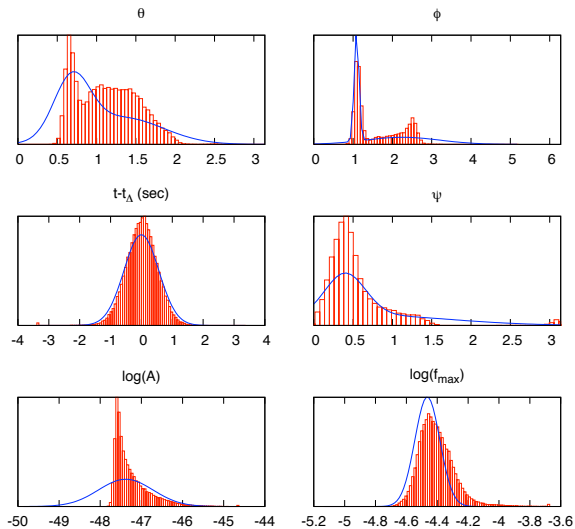} 
   \caption{The distribution of recovered parameter values for an MCMC search for MLDC training source 3.4.3 using simulated
LISA data. The Fisher Matrix Gaussian approximation is shown in blue for comparison.}
   \label{fig:hist3}
\end{figure}

An MCMC search for a signal in noisy data returns the PDF that maps out the uncertainty in each parameter.
This can be compared to the Gaussian approximation for the variance in each parameter derived from the
Fisher Information Matrix. The five distinct triggers from the search phase were used as starting points for
the MCMC runs. Data segments of length $16,384$ seconds, centered on the time of arrival of the burst, were used
in the second stage of the analysis. The parallel tempering routine used $N_C=12$ chains with a maximum
temperature of $T{\rm max} = 125$, spacing $\Delta T = 1.55$ and a $N=10^6$ iterations.
Histograms for the six source parameters for MLDC training source 3.4.3 are show
in Figure \ref{fig:hist3}.  The sky location parameters $\theta$ and $\phi$ have a bimodal distribution due to
the degeneracy of two sky locations from the detector symmetry.  The time of arrival of the signal at the detector
is very well determined, matching a Gaussian distribution with a variance of less than one second.  The maximum
frequency cutoff can also be well determined for sources such as this with $f_{\rm max}$ below the LISA transfer frequency.

\begin{figure}[htbp]
   \centering
   \includegraphics[width=3in]{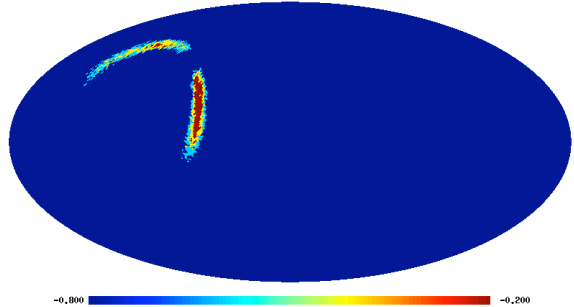}
   \caption{An MCMC search for an MLDC training data source with high SNR but a maximum frequency cutoff below the LISA transfer frequency.}
   \label{fig:source3}
\end{figure}

\begin{figure}[htbp]
   \centering
   \includegraphics[width=3in]{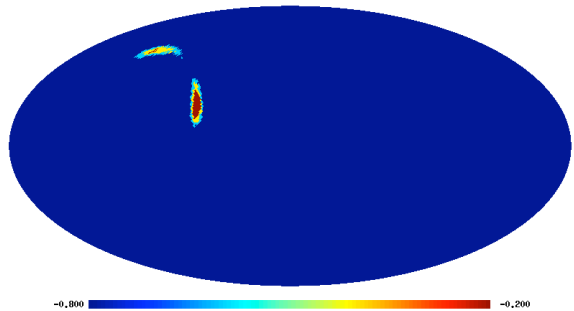}
   \caption{An MCMC search for a source with the same parameter values as Figure~\ref{fig:source3}, except for a
maximum frequency above the LISA transfer frequency.}
   \label{fig:source6}
\end{figure}

Although not much SNR is gained in the high frequency portion of the signal, the LISA transfer function
is essential for determining the direction to the source.  This is illustrated quite well by
training data sources 3.4.3 with SNR$\ =77$ and $f_{\rm max}=0.0115$ Hz.  Even this source with high SNR has poor angular
resolution (Figure \ref{fig:source3}).  A new search for a source with identical parameters, and almost identical SNR,
but with $f_{\rm max}=1.15$ Hz reveals the importance of the high frequency component of the signal (Figure \ref{fig:source6}).
	
\section{Advanced Ground Based Detectors}\label{aligo}

The wide frequency range of gravitational wave signals from cosmic string cusps allows them to be detected simultaneously
by space based and ground based gravitational wave detectors.  The current ground based detectors have not yet reported
a detection of a gravitational wave signal, but the next generation of detectors, such as Advanced LIGO, will have an
enhanced ability to detect gravitational waves, including bursts from cosmic string cusps. We investigated the performance
of the advanced terrestrial network by using the MLDC training source parameters, with the exception of the $f_{\rm max}$
parameter, which we set at 500 Hz to ensure that the signal will be in the LIGO-Virgo band. The relative orientation
of the ground and space based detectors was fixed by setting the start of the LISA observations to the autumnal equinox
in the year 2020.

When just considering the ground based network we use right ascension $\alpha$ and declination $\delta$ to
describe the sky location, and the polarization basis defined in Ref.~\cite{Anderson:2001if}. The time of arrival
of the bursts is referenced to the Earth geocenter. We use the reference Advanced LIGO wideband noise curve with strain spectral
density
\begin{equation}
S_n(f) = 10^{-49} \left(x^{-4.14} - \frac{5}{x^{2}} + 111 \left(\frac{2 -2 x^2 + x^4}{2 + x^2}\right) \right),
\end{equation}
where $x =f / 215 \, {\rm Hz}$.

\begin{figure}[htbp]
   \centering
   \includegraphics[width=3in]{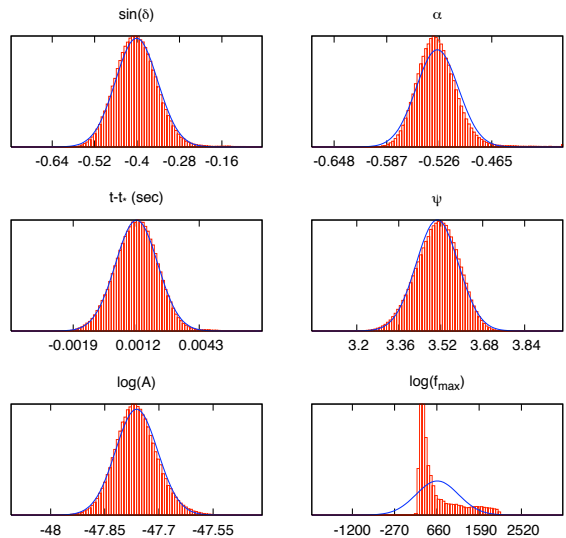} 
   \caption{The distribution of parameter values for an MCMC search of simulated data using the parameters for MLDC training source
3.4.0 (with $f_{\rm max}$ boosted to 500 Hz) for the advanced ground based detector network.  The Gaussian Fisher Matrix approximation is
included in blue for comparison. }
   \label{fig:LIGOhist0}
\end{figure}

\begin{figure}[htbp]
   \centering
   \includegraphics[width=3in]{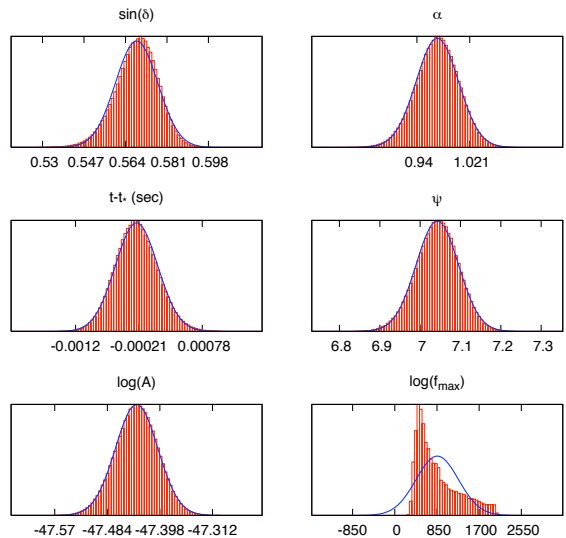} 
   \caption{The distribution of parameter values for an MCMC search of simulated data using the parameters for MLDC training source
3.4.3 (with $f_{\rm max}$ boosted to 500 Hz) for the advanced ground based detector network.  The Gaussian Fisher Matrix approximation is
included in blue for comparison. }
   \label{fig:LIGOhist3}
\end{figure}

The ground based detector network of the 4 km LIGO detectors in Hanford, WA and Livingston, LA and the Virgo detector
in Italy, if operating at advanced LIGO design sensitivity should respond to the MLDC sources with network SNRs of
24.3, 7.9, 4.3, 35.8, and 5.7 for sources $3.4.0 \rightarrow 3.4.4$, respectively. Without the benefit of joint LISA
observations, sources 3.4.2 and 3.4.4 may not be bright enough to be detected. Simulated data was generated for the LIGO-Virgo network
and single chain MCMC runs of $10^6$ iterations were used to produced the parameter histograms shown in
Figures \ref{fig:LIGOhist0} and \ref{fig:LIGOhist3}. The Fisher Matrix estimates for the parameter distributions are also
shown, and are seen to agree well with the MCMC derived PDFs. The $f_{\rm max}$ parameter is poorly determined,
resulting in a wide spread of $f_{\rm max}$ values visited by the MCMC chains. The sky location of the burst is determined
to an accuracy of $\sim 10$ square degrees.

\subsection{Combining Space and Ground Based Detectors}

The broad spectrum nature of signals from cosmic string cusps allows for joint observations between a space based detector
such as LISA and the network of ground based detectors.  The LISA sensitivity range is from 0.1 mHz to 1 Hz while LIGO is
sensitive in the range 10 Hz to 10 kHz.  A burst from a cosmic string cusp with a cut-off frequency above $\sim 50$ Hz
would produce a response across the LIGO-Virgo-LISA network, and the extra information from joint detection would improve
estimates of the source parameters.  One might expect that large Earth-LISA baseline (up to 170 seconds delay)
would result in extraordinary angular resolution, but the relatively poor determination of the time of arrival
at the LISA guiding center (of order 1 second) diminishes the effect.

For joint observations we adopted the Barycentric ecliptic coordinate system used to describe LISA observations. For the time
of arrival of the burst we used the geocenter time of arrival $(t_{\oplus})$, as this choice gives the smallest
correlation with the sky location uncertainty.

Table II shows the parameter estimation improvement gained by combining LISA with a ground based network detection. The gains
in sensitivity can mostly be attributed to the higher SNR. This can be understood by looking at the off-diagonal components
of the Fisher Information Matrix, and noting that even without the LISA contribution the parameter correlations are
already quite small. On the other hand, the addition of the LISA data reduces the sky location uncertainty to less than
a square degree, which improves the prospects of performing a microlensing follow-up.

\begin{table}[htbp]
   \centering
   \begin{tabular}{|c|c|c|c|c|}
      \toprule
      \multicolumn{2}{c}{} \\
      \hline
 MLDC 3.4.0 & $\ \ $ SNR  $\ \ $   & $ \sigma_{\theta} $ (rad)  &  $\sigma_{\phi} $ (rad)  & $\sigma_{t}$ (sec) \\ 
	\hline
	\hline
      \midrule
        
LIGO-Virgo & 24.3     & 5.7e-02 & 3.8e-02   & 1.2e-03            \\
LISA + LIGO-Virgo & 76.2      & 2.5e-02 & 9.6e-03   & 5.1e-04             \\     
 
 \hline
      \bottomrule
 \multicolumn{5}{c}{ }  \\
      \toprule

      \hline
 MLDC 3.4.1 & $\ \ $ SNR  $\ \ $   & $ \sigma_{\theta} $ (rad)  &  $\sigma_{\phi} $ (rad)  & $\sigma_{t}$ (sec) \\ 
	\hline
	\hline
      \midrule
        
LIGO-Virgo &  7.9             & 1.0e-02 & 6.9e-02   & 4.5e-04            \\
LISA + LIGO-Virgo &  22.9     & 4.7e-03 & 4.8e-02   & 2.4e-04             \\     
 
 \hline
      \bottomrule
 \multicolumn{5}{c}{ }  \\
      \toprule

      \hline
MLDC 3.4.3 & $\ \ $ SNR  $\ \ $   & $ \sigma_{\theta} $ (rad)  &  $\sigma_{\phi} $ (rad)  & $\sigma_{t}$ (sec) \\ 
	\hline
	\hline
      \midrule
        
LIGO-Virgo & 35.8      & 1.7e-02 & 2.4e-02   & 3.5e-04            \\
LISA + LIGO-Virgo & 84.8     & 5.3e-03 & 2.4e-03   & 5.9e-05             \\     
 
 \hline
      \bottomrule
   \end{tabular}
   \caption{A joint detection by LISA and an advanced ground based network results in improvements in parameter estimation
as shown for three sources from the MLDC 3.4 training data.}
   \label{tab:joint2}
\end{table}
	
\section{Conclusion}\label{con}

A matched filter analysis using parallel tempered Markov chain Monte Carlo techniques can both detect and characterize
the gravitational wave signals from cosmic string cusps in simulated LISA and LIGO-Virgo data.  Cosmic string cusp waveforms
have a broad enough frequency range that they can be detected jointly by space and ground based detectors, and these joint
detections significantly improve the angular resolution over what can be achieved by LISA alone. The addition of LISA
observations to the ground based network leads to a more modest improvement that may improve the prospects of finding
electromagnetic counterparts from microlensing of stars in foreground galaxies.

The analysis presented here can be improved in several ways. Rather than performing the search with the SNR of the best fit
template being used as a frequentist detection statistic, the Bayesian evidence for the signal can be computed by thermodynamic
integration across the parallel chains~\cite{Goggans:2004aa}. The instrument noise levels must be treated as search parameters
when computing the evidence, and the noise model should take into account the non-symmetric noise levels found in the
Challenge 3.4 data. The signal model can be extended to account for multiple overlapping bursts, which would be an improvement
on the sequential regression used in the current analysis. A more realistic analysis should also take into account the
confusion background that would accompany the bright burst signals.\\

\section{Acknowledgements}	

This work was supported by NASA grant NNX07AJ61G.
The full sky maps generated for this paper have been produced using the HEALPix (G\'orski et al., 2005~\cite{Gorski:2005rw}) package.
We thank Tyson Littenberg for helpful discussion about the parallel tempered MCMC algorithm, and Xavier Siemens and Ray Frey for
providing comments on an earlier draft of the paper.

\bibliography{strings}

\end{document}